\documentclass[11pt,a4paper]{article}
\usepackage{graphicx}
\usepackage{color}

\usepackage{amsmath}
\usepackage{amssymb}

\usepackage{natbib}

\newcommand{\mylab}[3]{\raisebox{#2}[0mm][0mm]{\makebox[0mm][l]{\hspace*{#1}\textbf{#3}}}}

\definecolor{rred}{rgb}{0.8, 0, 0}
\definecolor{bblue}{rgb}{0, 0, 0.8}
\definecolor{ggreen}{rgb}{0, 0.5, 0}

\newcommand{\opentriangle}{\mbox{$\triangle$}}
\newcommand{\opentriangledown}{\mbox{$\bigtriangledown$}}

\newcommand{\dashed}{\protect\mbox{-\; -\; -\; -}}

\begin{document}

\title{Wall-bounded turbulence control: statistical characterisation of actions/states}

\author{R Pastor$^1$, A Vela-Mart\'{\i}n$^2$ 
and O Flores$^1$ \\ 
$^1$ Dept. Bioingenier\'ia e Ing. Aeroespacial, Universidad Carlos III de Madrid, Spain. \\ 
$^2$ School of Aeronautics, Universidad Polit\'ecnica de Madrid, Spain.}

\maketitle

\begin{abstract}
The present paper reports the results of a Monte Carlo experiment using a turbulent channel flow. 
Different actions are proposed, varying the size, duration and sign of a localised volumetric force that acts near one wall of a turbulent channel flow, running at a small Reynolds ($Re_\tau=165$) in a small computational domain. 
The effect of each action is evaluated comparing the evolution of the flow with and without the action, gathering statistics over 1700 repetitions of the experiment for each action. 
The analysis of the results show that small/short forcings are equally likely to increase or decrease the skin friction drag, independently of the sign of the forcing (i.e., towards or away from the wall). When the size or the duration of the forcing increases, so does the probability of increasing the skin friction. 
Then, an {\em a priori} analysis is performed, evaluating the state of the flow just before the action, conditioned to actions that result in a decrease of the skin friction over a period of one eddy turn-over time.  
The resulting fields of velocity, wall shear stresses and wall pressure are consistent with an opposition control strategy, where the forcing is opposing the vertical motions near the wall. 
Finally, a preliminary analysis of the performance of actuation triggered by pressure or wall shear stresses sensors is evaluated (i.e., {\em a posteriori} analysis). 
Our results show that the actuation triggered by a wall shear sensor seems to be more effective than the actuation triggered by a wall pressure sensor, at least for the preliminary definitions of sensors and thresholds used here. 
\end{abstract}

\section{Introduction}

In many industrial problems, engineers face the challenge of controlling the flow in a turbulent boundary layer developing over a solid surface or wall.  Given the role that turbulent motions play in the transport of momentum, heat and mass, the typical objective of this control is either to damp the turbulent motions (i.e., to reduce the skin friction drag or prevent heat transfer), or to promote them (i.e., to increase mixing or heat transfer). In most applications of practical interest this control needs to be done using actuators and sensors that are placed at the wall. This requires establishing which turbulent motions can be observed from the wall, and which turbulent motions can be controlled from the wall. Moreover, for those that can be observed and controlled, it is necessary to evaluate as well what type of actuation is needed, both in space and time \citep{rowley2017model}.

Recent research efforts have focused on linking the prevalent structures identified in wall bounded turbulent flows \citep{adrian2007hairpins, delalamo2006self, kawahara2012significance, jimenez2012cascades} to variables that can be measured at the wall, like the pressure or the shear stresses at the walls. For instance, the analysis of the cross-correlation between the wall pressure and the wall-normal velocity fluctuations \citep{sanmiguel2018wall} shows that the spatially-averaged wall pressure over a length $L$ is correlated with vertical velocity structures at a wall distance $y$, which increases with $L$. More recently, 
Encinar et al  \citep{encinar2018CTR} have evaluated the ability to estimate the velocity fluctuations far from the wall using measurements of pressure and shear stresses at the wall. Both studies suggest that sensing wall-attached flow structures is feasible, but that our ability to estimate their detailed structure far from the wall is limited. 

In terms of how to apply the control, the most common control strategy employed in wall-bounded turbulence for drag reduction is the so called {\em opposition control} where blowing and suction at the wall is applied in opposition to the vertical velocity fluctuations sufficiently close to the wall \citep{choi1994active,abbassi2017skin}. 
Recent attempts at opposition control \citep{yun2018opposition,park2018prediction} use neural networks and deep-learning techniques to estimate the near-wall vertical velocity, and hence determine the control, from the pressure at the wall. 


It should be noted that, while these studies show various degrees of turbulence damping and/or skin friction reduction,  it is not completely clear whether this strategy is practical in a real world application. The required sensor or actuators might become too small to control the near-wall region streaks. Also, it might be possible that it is not necessary to oppose the turbulent motions at all times, or that controlling at certain times is just too ineffective. Another problem in some of those works is the fact that either the control, or the sensor (or both) involve information on the whole wall, which will be unfeasible in a real world application, where the sensor and the actuator will be local in space.  

In order to investigate the issue of the locality of sensors and actuators, both in space and time,  the present project proposes a statistical evaluation of the {\em effect} produced by a localised {\em action} applied to a turbulent flow in a channel, when the latter is in a certain {\em state}. This will be done by performing a Monte Carlo experiment, where a large number of realisations (i.e., episodes) of the effect that each action produces on the flow will be simulated. 
Then, an {\em a priori} analysis will be performed, where the specific states that result in the desired effect (i.e, skin friction reduction) will be identified for each action. This information will be used to define a preliminar localised sensor at the wall, which will be used in an {\em a posteriori} analysis where the net effect of an action triggered by this localised sensor will be evaluated. 

The present work should be understood as a proof-of-concept test,  to evaluate the capabilities of this Monte Carlo experiment. As such, the selected turbulent channel configuration is a toy problem: a low-Reynolds number case in a small computational domain, which can run in very short times in a GPU.  

The rest of the paper is organised as follows. First, the numerical simulations and the Monte Carlo experiment are described. 
Then, section 3 presents the results of the experiment, starting with a statistical characterisation of the change in the skin friction produced by the actions,  without taking into account the state of the flow at the time of actuation. 
Subsection 3.1 presents the {\em a priori}  analysis, and subsection 3.2 presents the definition of the localised sensor and the {\em a posteriori} analysis. 
The article finishes with some conclusions and guidelines for future work. 

\section{Numerical experiment}

The toy problem considered for the present work is an incompressible turbulent flow in a plane channel, driven by a constant mass flux. 
The friction Reynolds number is $Re_\tau = u_\tau h/\nu = 165$, where $u_\tau$ is the friction velocity,  $h$ is the channel half-height and $\nu$ is the kinematic viscosity. The simulations are run in a small computational domain, with streamwise and spanwise periodicities equal to $L_x = \pi h$ and $L_z = \pi h/2$, respectively. The computational domain and Reynolds number result in a turbulent channel were the buffer region and the outer region have roughly the same scale. However, this small channel still provides a good representation of the structures and the dynamics of the near-wall region \citep{jimenez1991minimal,flores2010hierarchy}. 

The channel simulations are run using a GPU enabled version of a channel code \citep{encinar2018second}. The code uses a formulation of the Navier-Stokes equations based on the wall-normal component of the vorticity and on the Laplacion of the wall-normal velocity, which are integrated with a semi-implicit third-order low-storage Runge-Kutta. The spatial discretisation is based on a Fourier expansion on the wall-parallel directions ($x$ and $z$), using a pseudo-spectral method for the computation of the non-linear terms. In the wall normal direction, a non-uniform grid and a seven-point compact finite differences with spectral-like resolution are used.  The spatial discretisation uses  $N_x = N_z = 32$ and $N_y = 128$ grid points, yielding a resolution in the horizontal directions of $\Delta x ^+ = 8$ and $\Delta z^+ = 4$ before dealiasing. In the vertical direction, the non-uniform grid provides $\Delta y^+_{\min} = 0.4$ at the walls, and $\Delta y^+_{\max} = 5.4$. at the centre of the channel. 
For convenience, the wall at $y=0$ is denoted ``bottom wall'', and the wall at $y=2h$ is the ``upper wall'', even if there are no buoyancy effects in the simulations.  

The Monte Carlo experiment is designed as follows. First, the base channel configuration (without forcing) is run until a statistically steady state is obtained. Then, taking a snapshot at $t=t_0$ as initial condition, the base channel configuration is run from $t=t_0$ to $t=t_0 + T_{epi}$, where $T_{epi}$ defines the duration of each episode. For each actuation (defined below), a simulation using the same initial condition is run for the same time interval, allowing a direct evaluation of the effect of the actuation by comparing the actuated time series with the base channel simulation. This process is repeated for a number of episodes $N_{epi}$, using as initial condition for episode $n$ the last snapshot of the base channel run in episode $n-1$. 

The duration of the episodes $T_{epi}$ is chosen long enough to ensure that the initial conditions of consecutive episodes are statistically independent. 
This is evaluated with the autocorrelation of the instantaneous skin friction (not shown), where the instantaneous skin friction of the bottom wall is defined as
\begin{equation} 
\label{eq:tau}
\tau(t) = \frac{1}{L_x L_z} \int_0^{L_x} \!\!\int_0^{L_z} \mu \frac{\partial u}{\partial  y}(x,0,z, t) dx dz. 
\end{equation} 
Preliminary results show that the autocorrelation has a negative peak at $t \approx 2 h/u_\tau$, and its absolute value becomes smaller than 0.01 at $t \gtrsim 3 h/u_\tau$. Hence,  the duration of the episodes is chosen to be $T_{epi} = 3.6 h/u_\tau$.

The actions considered in this work corresponds to a volumetric force applied for a short time ($T_f$) in the vertical direction. This force acts in a volume close to the wall defined by a Gaussian with a characteristic size $L_f$, 
\begin{equation} 
f_y(x,y,z,t) = \left\{ \begin{array}{ll}
f_0 \frac{y}{h} \exp\left( \frac{-4 (x^2 + z^2 + y^2) }{L_f^2} \right) & \mbox{if $t_0 \le t \le T_f$} \\ 
0 & \mbox{else,} 
\end{array} \right.
\label{eq:forc}
\end{equation}
where $y$ is the vertical distance to the lower wall of the channel. 
The vertical profile of the forcing is such that $f_y=0$ at the wall, peaking at $y/L_f \approx 0.35$ and becoming negligible for $y\gtrsim L_f$. 
After some preliminary tests, 12 different actions are considered in this study, by varying the size ($L_f^+ = 50, 100$ ), duration ($T_f^+ = 25, 50, 100$) and intensity ($f_0 = \pm 8 u_\tau^2/h$) of the vertical force. The number of episodes is $N_{epi} = 17000$ for all actions, except for those with $T_f^+ = 100$ which have $N_{epi} = 2145$. 

To some extent, the Monte Carlo experiment proposed here is similar to the experiment proposed by Jimenez \cite{jimenez2018machine}. In the latter, the a 2D isotropic turbulent flow was {\em interrogated} by removing pieces of the fluid (i.e., setting the velocity or vorticity equal to zero), and checking its effect on the evolution of the kinetic energy and the total enstrophy. Here, instead of removing pieces of the fluid, a volumetric force is applied close to the wall. And the effect on the flow is measured in terms of the evolution of the skin friction, rather than the kinetic energy or the enstrophy of the flow. 


Two different averages are used in the analysis present in section 3. Ensemble averages over the $N_{epi}$ episodes  are denoted with $\langle \rangle$. The time average of a variable $\phi(t)$ over a time $T$ is denoted with an overline, 
\begin{equation} 
\label{eq:aver}
\overline{\phi}(T) = \frac{1}{T} \int_0^T \phi(t) dt.
\end{equation}
 
\section{Results} 

\begin{figure}
\includegraphics[width=0.47\textwidth]{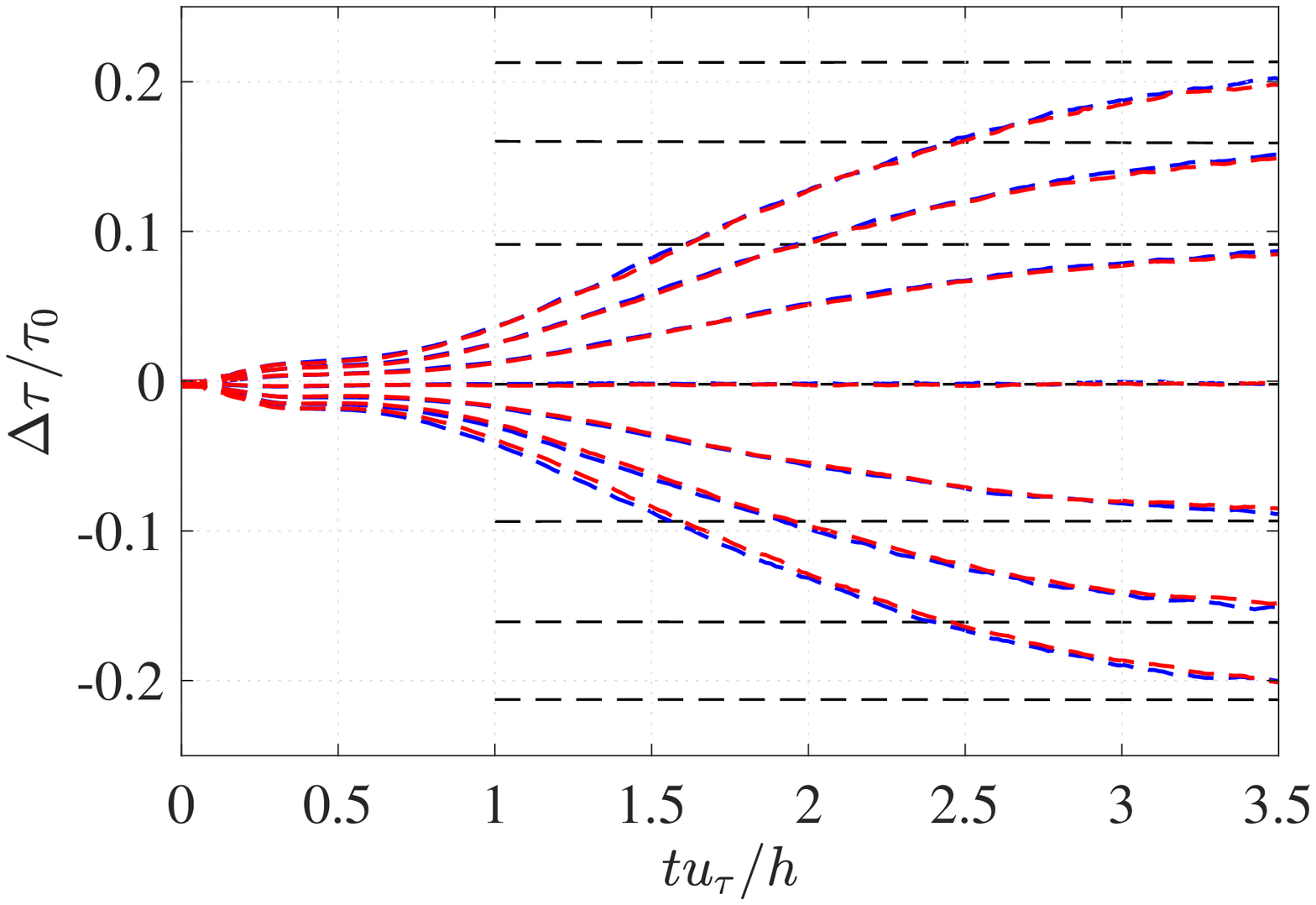}
\mylab{-6.3cm}{4.7cm}{$(a)$}
\hfill
\includegraphics[width=0.47\textwidth]{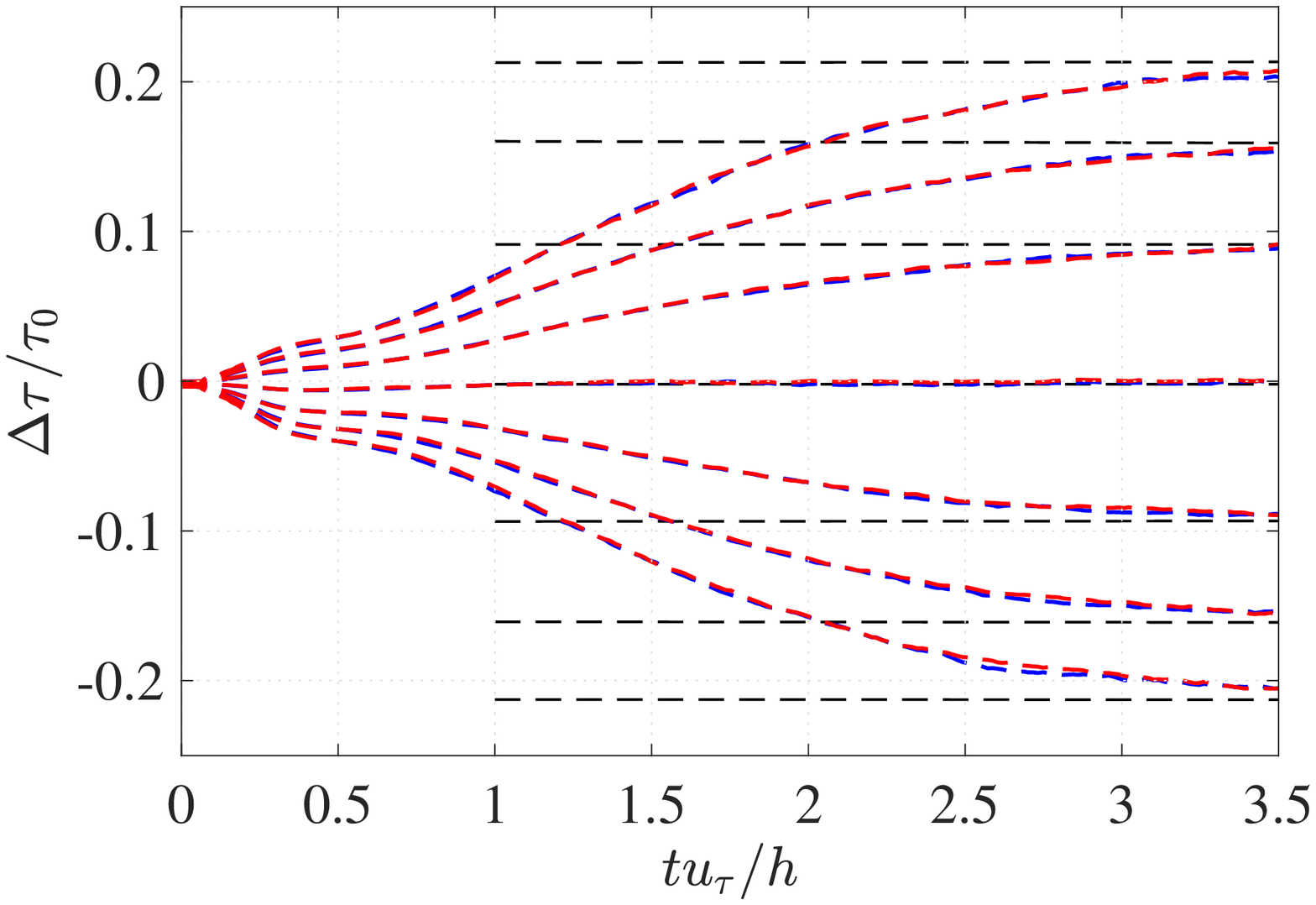}
\mylab{-6.3cm}{4.7cm}{$(b)$}
\vspace{2mm}

\includegraphics[width=0.47\textwidth]{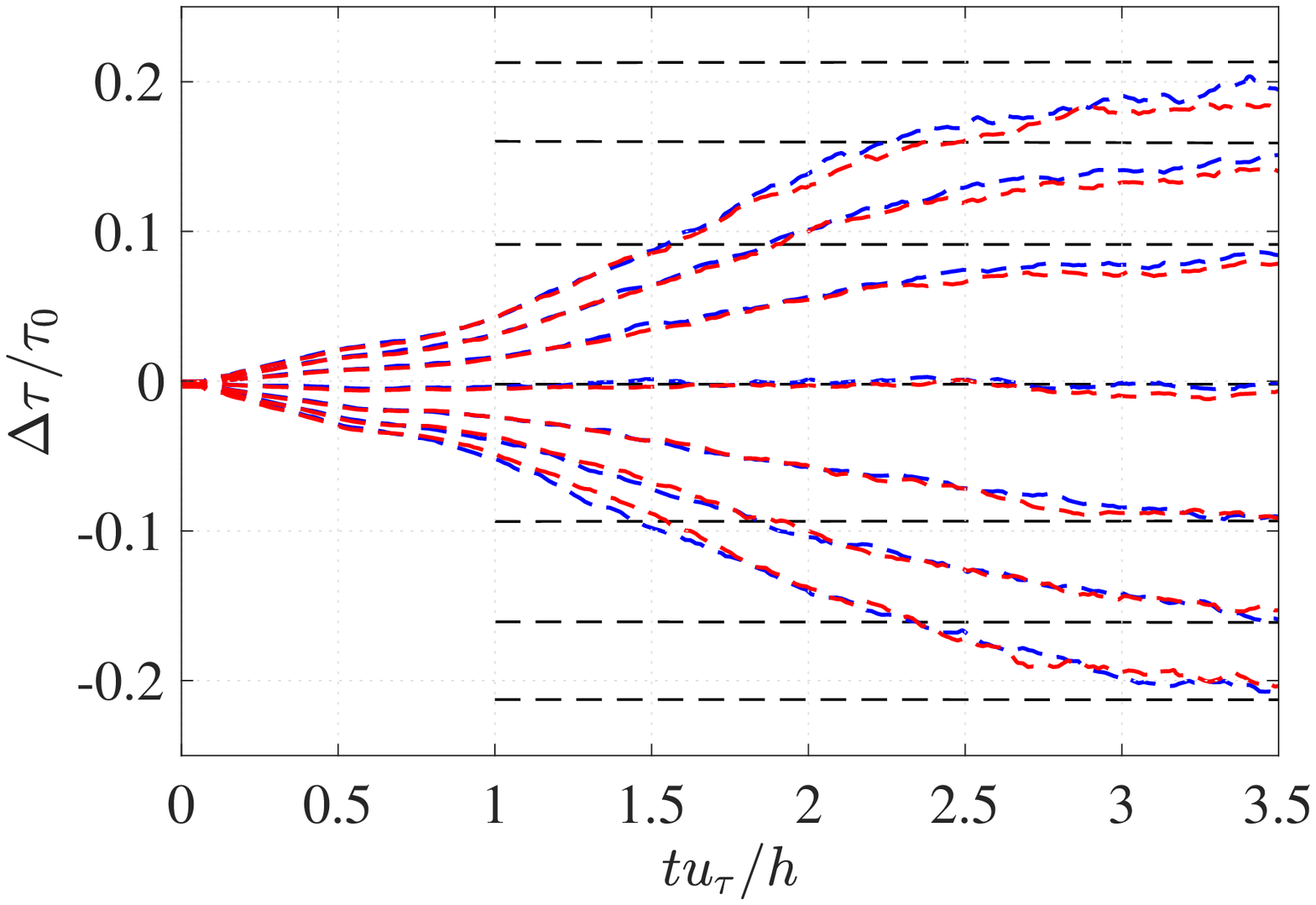}
\mylab{-6.3cm}{4.7cm}{$(c)$}
\hfill
\includegraphics[width=0.47\textwidth]{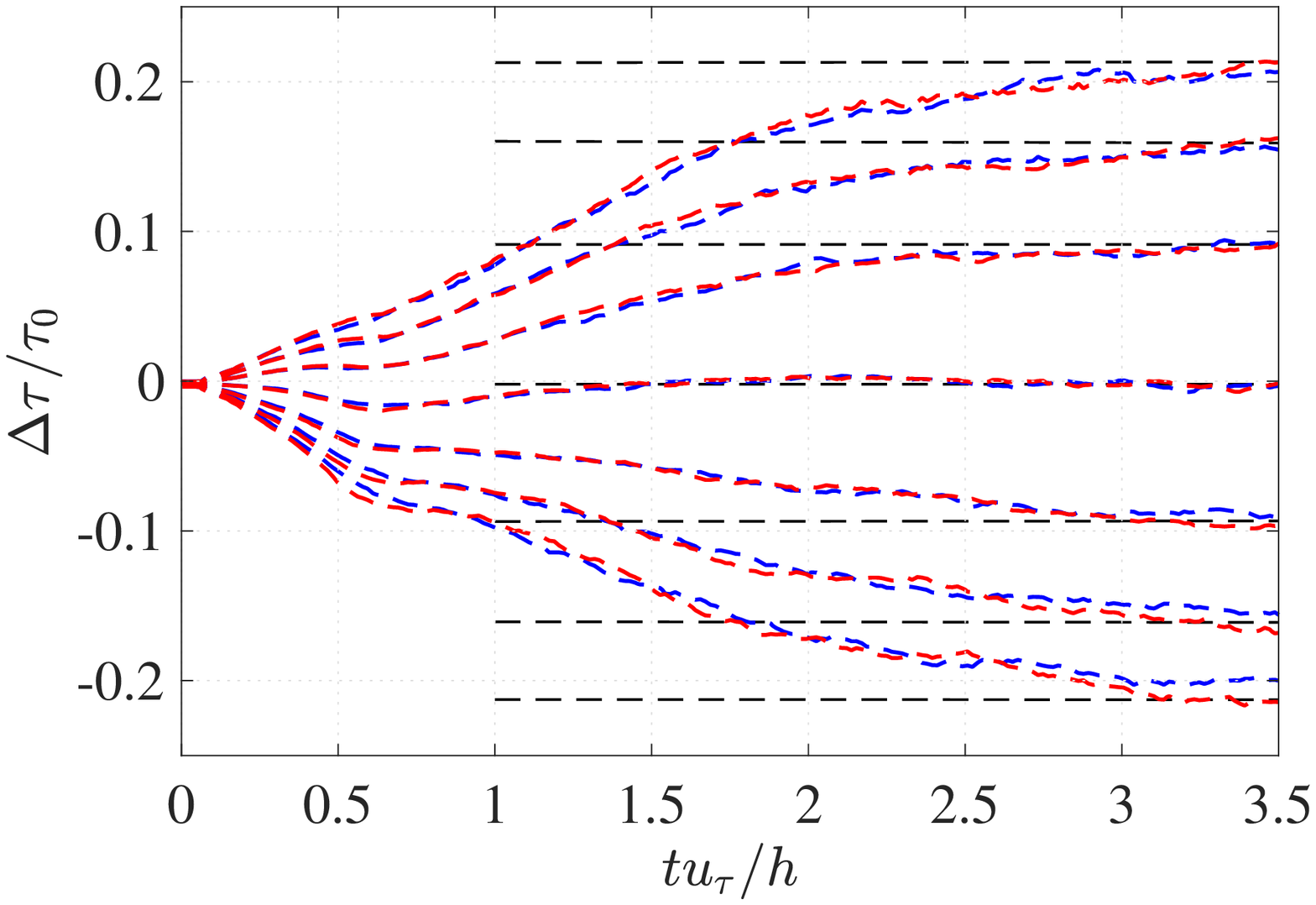}
\mylab{-6.3cm}{4.7cm}{$(d)$}
\caption{
\label{fig:cdf_all} 
Cumulative Distribution Function (CDF) for $\Delta \tau / \tau_0$ versus time, where $\tau_0$ is the mean wall shear of the base flow. 
$\color{blue}\dashed$ for $f_0 = + 8 u_\tau^2/h$ (i.e., upward). 
$\color{red}\dashed$ for $f_0 = - 8 u_\tau^2/h$ (i.e., downward). 
($a$) $L_f^+ = 50$ and $T_f^+ = 50$. 
($b$) $L_f^+ = 100$ and $T_f^+ = 50$.  
($c$)  $L_f^+ = 50$ and $T_f^+ = 100$. 
($d$)  $L_f^+ = 100$ and $T_f^+ = 100$. 
In all panels, the contours of the CDF correspond to $[0.06, 0.12, 0.25, 0.5, 0.75, 0.88, 0.94]$,  and the horizontal black dashed lines show data from the CDF of $\tau(t) - \tau(t+4h/u_\tau)$ for the base channel. 
}
\end{figure}

In order to assess the effect of the actions on the channel, figure \ref{fig:cdf_all} shows the Cumulative Distribution Function (CDF) of the difference between the instantaneous skin friction in the lower wall of the base channel and the forced channels, 
$\Delta \tau = \tau_{base}(t) - \tau_{forced}(t)$, where $\tau(t)$ is defined above in equation (\ref{eq:tau}). Note that a positive value of $\Delta \tau$ corresponds to an action that results in drag decrease, while a negative value corresponds to drag increase. Also, only the skin friction at the bottom wall is considered, since the volumetric force defined in (\ref{eq:forc}) is only applied on that side of the channel. For reference, the change in wall shear $\Delta \tau$ is normalised with the skin friction of the base case, $\tau_0 = \lim_{T\to\infty} \overline{\tau}(T)$. 

Figure \ref{fig:cdf_all}$a$ shows the result for the case with $L_f^+ = 50$ and $T_f^+ = 50$. The time evolution of the CDF suggests three different stages. For $t\lesssim T_f \approx 0.3h/u_\tau$, the direct effect of the forcing starts producing differences between the evolution of the skin friction of the base and forced cases. For $ t \gtrsim h/u_\tau$, the width of the CDF increases monotonously with $t$ until the contours reach the same values obtained for the CDF of the difference between the skin friction of two base channels at uncorrelated times (i.e., the CDF of $\tau_{base}(t) - \tau_{base}(t+4h/u_\tau)$, indicated in the figure by black horizontal lines). In between, for $0.3h/u_\tau \lesssim t \lesssim h/u_\tau$ the CDF varies slowly with time, suggesting that 
for those times the 
perturbation generated by the action still dominates over the natural perturbations developing in the turbulent flow. 
It is very interesting to note that the CDF is symmetric for all times, suggesting that the actuation is equally likely to produce drag increase ($\Delta \tau < 0$) or decrease ($\Delta \tau >0$). Also, there are no difference in the CDF of the actions with positive and negative $f_0$, suggesting that the same effect can be achieved forcing the flow upwards and downwards. 

Figures \ref{fig:cdf_all}$b$ to $d$ show the effect on $\Delta \tau$ of increasing the size and the duration of the volumetric forcing. When the size of the forcing is increased (compare figure \ref{fig:cdf_all}$a$ to $b$), the spread in $\Delta \tau$ becomes larger, consistent with a stronger effect on the flow. 
When the forcing acts for a longer time (compare figure \ref{fig:cdf_all}$a$ to $c$), the initial stage becomes longer, yielding a stronger net effect on the skin friction. Interestingly, for the case with the largest and longest forcing (figure \ref{fig:cdf_all}$d$, $L_f^+ = 100$ and $T_f^+=100$) we observe that the median value of $\Delta \tau$ becomes negative. This implies that this forcing is consistently modifying the structure of the near-wall region in a way that increases the skin friction, both for positive and negative values of $f_0$. It should be noted that the flow through time for this toy problem is relatively short: assuming a convection velocity of $10u_\tau$ \citep{del2009estimation}, the flow-through time is about $50 \nu/u_\tau$. Hence, when $T_f^+ = 100$, every fluid particle aligned with the actuator flows over it twice per episode. From this point of view, only the actions with $T_f^+ = 25$ are strictly local in the streamwise direction for the present computational domain. 

\begin{figure}
\includegraphics[width=0.47\textwidth]{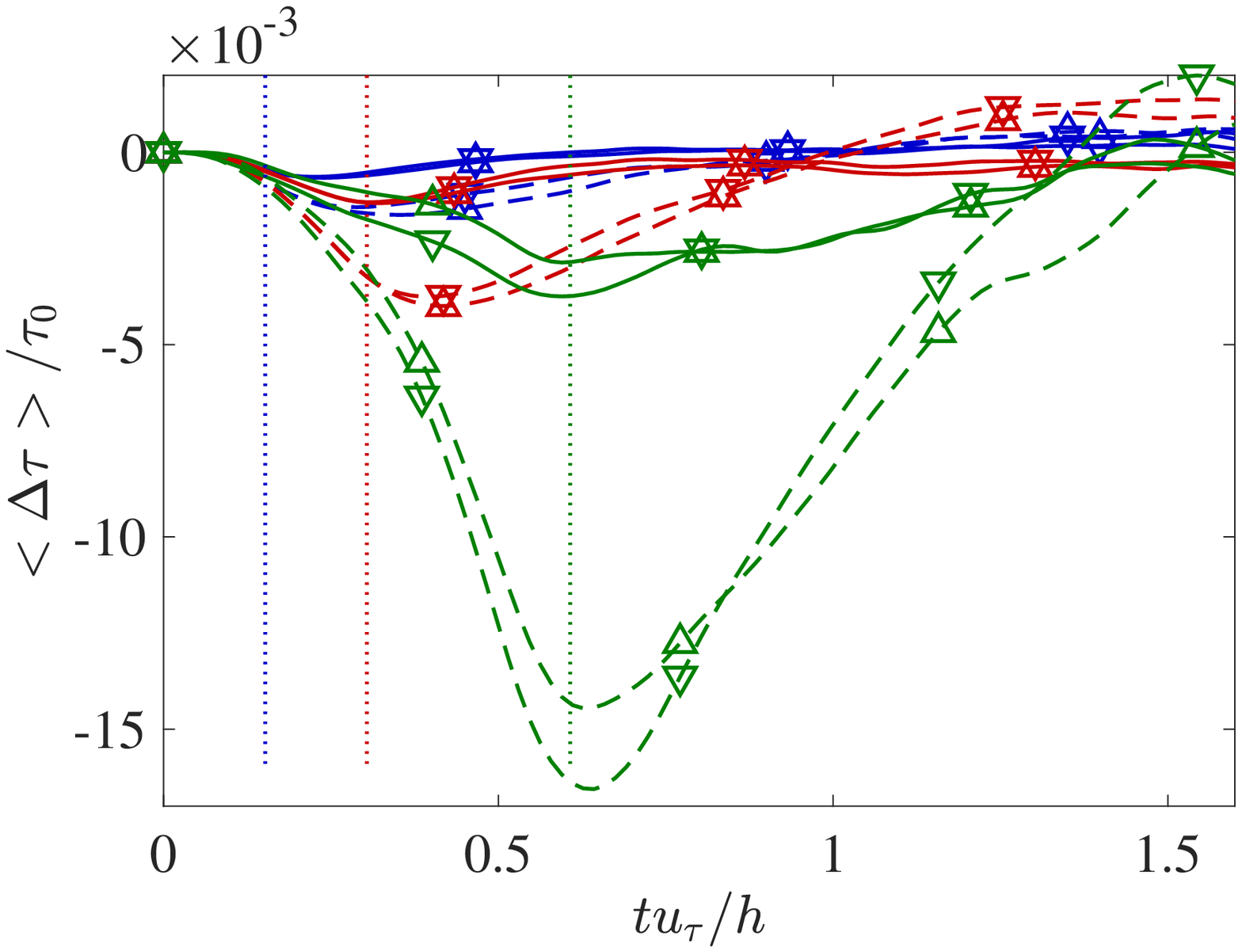}
\mylab{-1.cm}{1.3cm}{$(a)$}
\hfill
\includegraphics[width=0.47\textwidth]{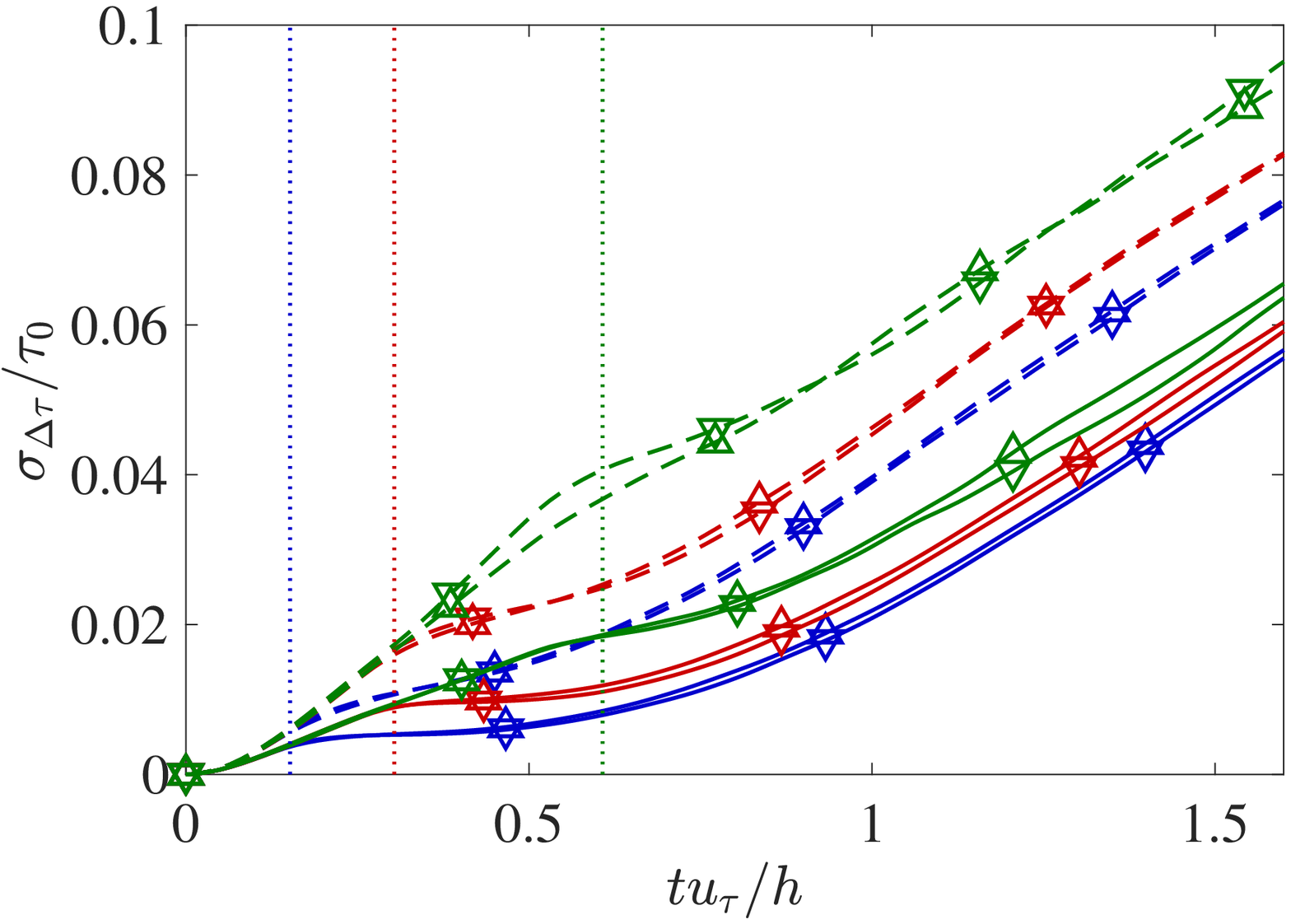}
\mylab{-1cm}{1.3cm}{$(b)$}
\caption{
\label{fig:mean_std} 
($a$) Mean and ($b$) standard deviation of $\Delta \tau$ versus time, normalised with the mean skin friction of the base flow. 
Lines and symbols indicate the characteristics of the forcing: 
$\opentriangle$ for $f_0 = +8 u_\tau^2/h$, $\opentriangledown$ for $f_0 = -8 u_\tau^2/h$; 
solid lines for $L_f^+ = 50$, dashed for $L_f^+ = 100$; 
{\color{bblue} blue} for $T_f^+=25$, {\color{rred} red} for $T_f^+ = 50$ and {\color{ggreen} green} for $T_f^+ = 100$. 
In both panels, the dotted vertical lines indicate $t=T_f$.  
}
\end{figure}   

The trends observed in figure \ref{fig:cdf_all} are more clearly observed in figure \ref{fig:mean_std}, where the ensemble mean and standard deviation of $\Delta \tau$ are represented as a function of time. The tendency of the forcing to shift the CDF towards the drag increase (i.e., $\Delta \tau<0$) is visible in the mean value for all cases in figure \ref{fig:mean_std}$a$, not only on the strongest and longest forcing. In any case, the shift in the mean value is always small, below 1.5\% of $\tau_0$ for the strongest forcing, and smaller than 0.5\% for the rest of the actuations. The three different stages of the time evolution of the CDF discussed above are clearly observed in the evolution of the standard deviation in figure \ref{fig:mean_std}$b$. All cases show the same relatively quick growth of $\sigma$ for $t\gtrsim h/u_\tau$, suggesting that for those times the 
natural perturbations developing in the turbulent flow are overriding the perturbations directly introduced by the actuation.


\subsection{{\em A priori} analysis: identification of the states where actuation results in skin friction reduction} 

In order to identify the states where a given action results in drag reduction, it is necessary first to provide a quantitative definition of drag reduction \footnote{
The definition of drag reduction used here requires having less skin friction than the base configuration instantaneously. It could also be possible to define drag reduction as having less skin friction than the averaged value of the base case. 
}. Given the results shown in figures \ref{fig:cdf_all} and \ref{fig:mean_std}, it seems sensible to use for that purpose $\overline{\Delta \tau}(h/u_{\tau})$, the time averaged value over the first eddy turnover of the difference in wall shear between the base case and the forced case (see equation \ref{eq:aver}).

Only the cases with $L_f^+=100$ and $T_f^+ = 25$ will be considered in this section. The cases with the smaller forcing ($L_f^+=50$) have weaker effects on the flow, and qualitatively behave similarly to cases with $L_f^+=100$. Nonetheless, the present toy problem basically has only one length-scale, since at $Re_\tau \approx 160$ the size of the velocity structures of the near-wall region is comparable to the height of the channel.  
The cases with forcings longer in time are more strongly biased towards drag increase, and they could be argued to not represent a completely local actuation on the flow, as discussed above in terms of the flow-through times. In any case, the results presented here for  $L_f^+=100$ and $T_f^+ = 25$ are very similar to those obtained for the other forcings.

\begin{figure}
\includegraphics[width=0.47\textwidth]{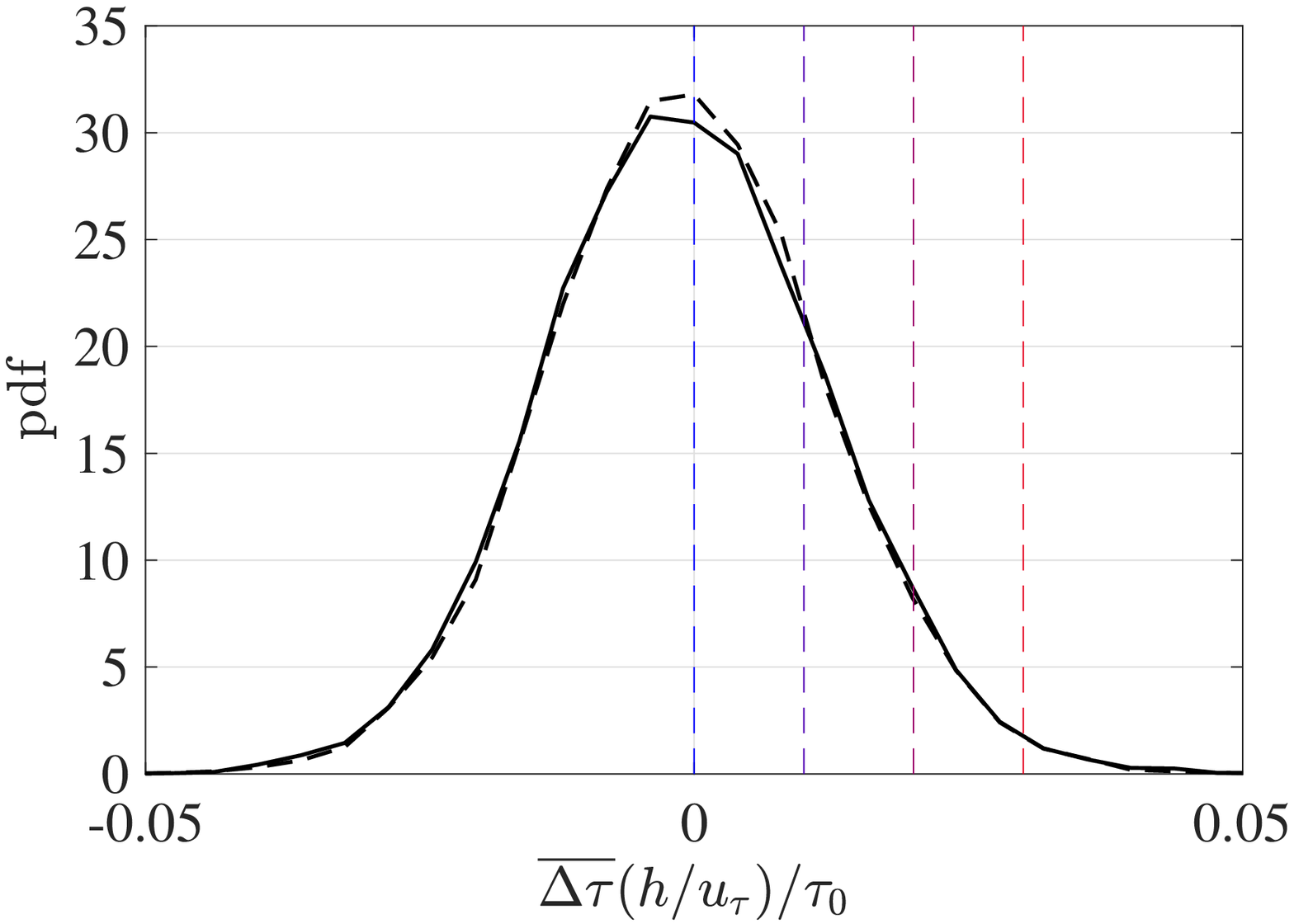}
\mylab{-6.6cm}{4.7cm}{$(a)$}
\hfill
\includegraphics[width=0.47\textwidth]{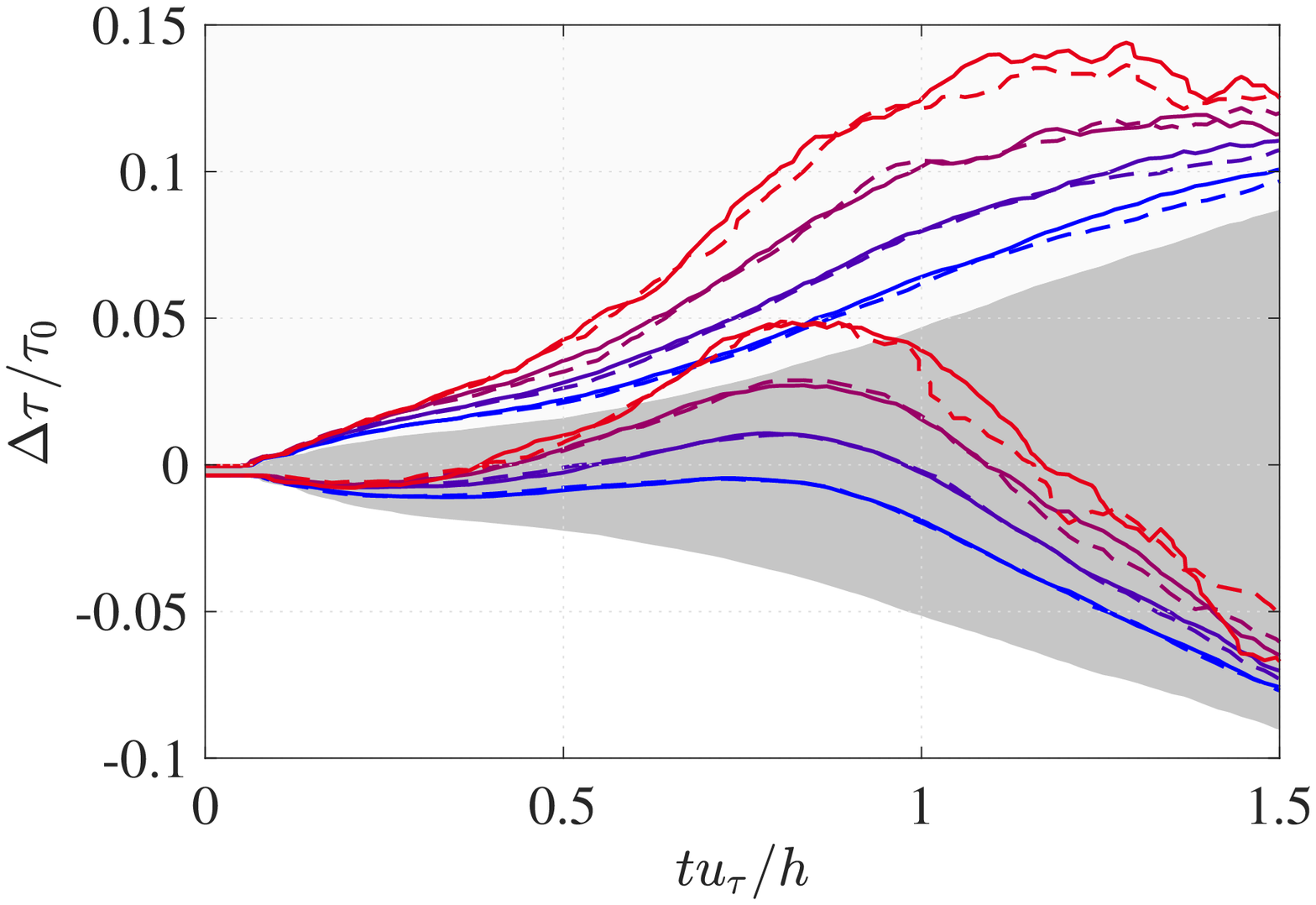}
\mylab{-6.3cm}{4.7cm}{$(b)$}
\caption{
\label{fig:cdf_c} 
($a$) Probability Density Function (PDF) of $\overline \Delta \tau (h/u_{\tau})$  and 
($b$) cumulative distribution functions of $\Delta \tau / \tau_0$ conditioned to $\overline \Delta \tau (h/u_{\tau}) > \alpha \tau_0$. 
Both plots correspond to case $L_f^+= 100$ and $T_f^+=25$, solid lines for  $f_0>0$ and dashed lines for $f_0<0$. 
In ($b$), the shaded patch corresponds to the un-conditioned CFD (same as plotted in figure \ref{fig:cdf_all}), and the lines correspond to the conditioned CDF for various $\alpha$, from $\alpha=0$ (blue) to $\alpha = 0.03$ (red).  
In all cases, the plotted contours are 0.1 and 0.9. 
For convenience, the thresholds used in ($b$) are also indicated in ($a$) with vertical lines with the same color code. 
}
\end{figure}   

In order to define a proper threshold for $\overline{\Delta \tau}(h/u_\tau)$, figure \ref{fig:cdf_c}$a$ shows its Probability Density Function (PDF) for case $L_f^+= 100$ and $T_f^+=25$.  Based on the range of the PDF, four thresholds are defined to condition $\Delta \tau$, as indicated by the vertical lines in the figure. Then, the CDF of $\Delta \tau$ is conditioned to $\overline{\Delta \tau}(h/u_\tau)$ larger than those thresholds, as shown in figure \ref{fig:cdf_c}$b$. It can be observed that the conditioned CDF is shifted towards positive values (drag reduction with respect to the unforced case), and that the shift becomes larger as the value of the threshold increase. As before, there is no difference between the positive and negative forcings (i.e., upward and downward forces). For times longer than $h/u_\tau$ the conditional CDF gradually tends towards the unconditioned CDF, and for $t \gtrsim 2h/u_\tau$ (not shown) there is no discernible difference between both CDFs. 

\begin{figure}
\includegraphics[width=0.32\textwidth]{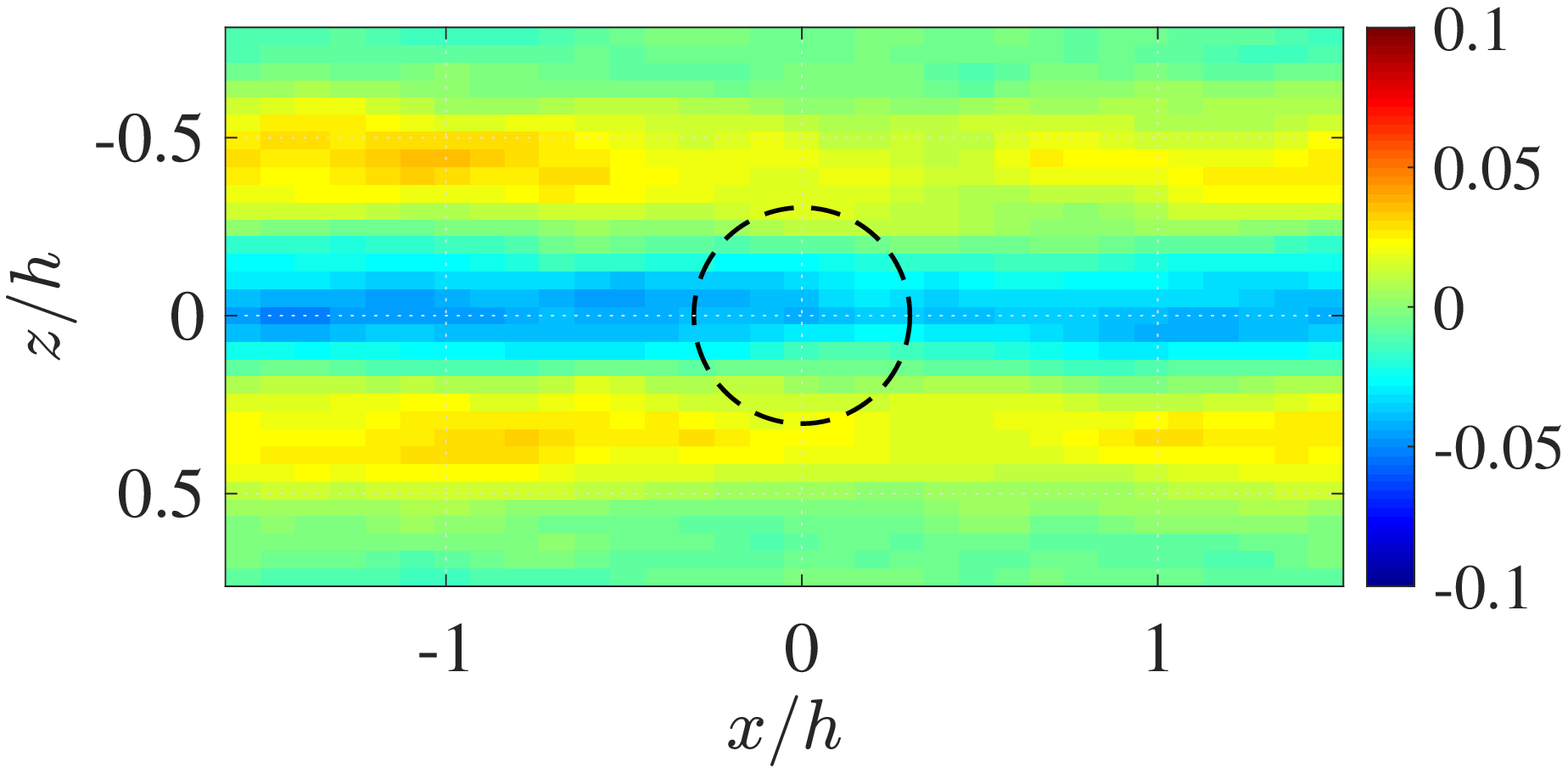}
\mylab{-5.3cm}{2.5cm}{$(a)$}
\hfill
\includegraphics[width=0.32\textwidth]{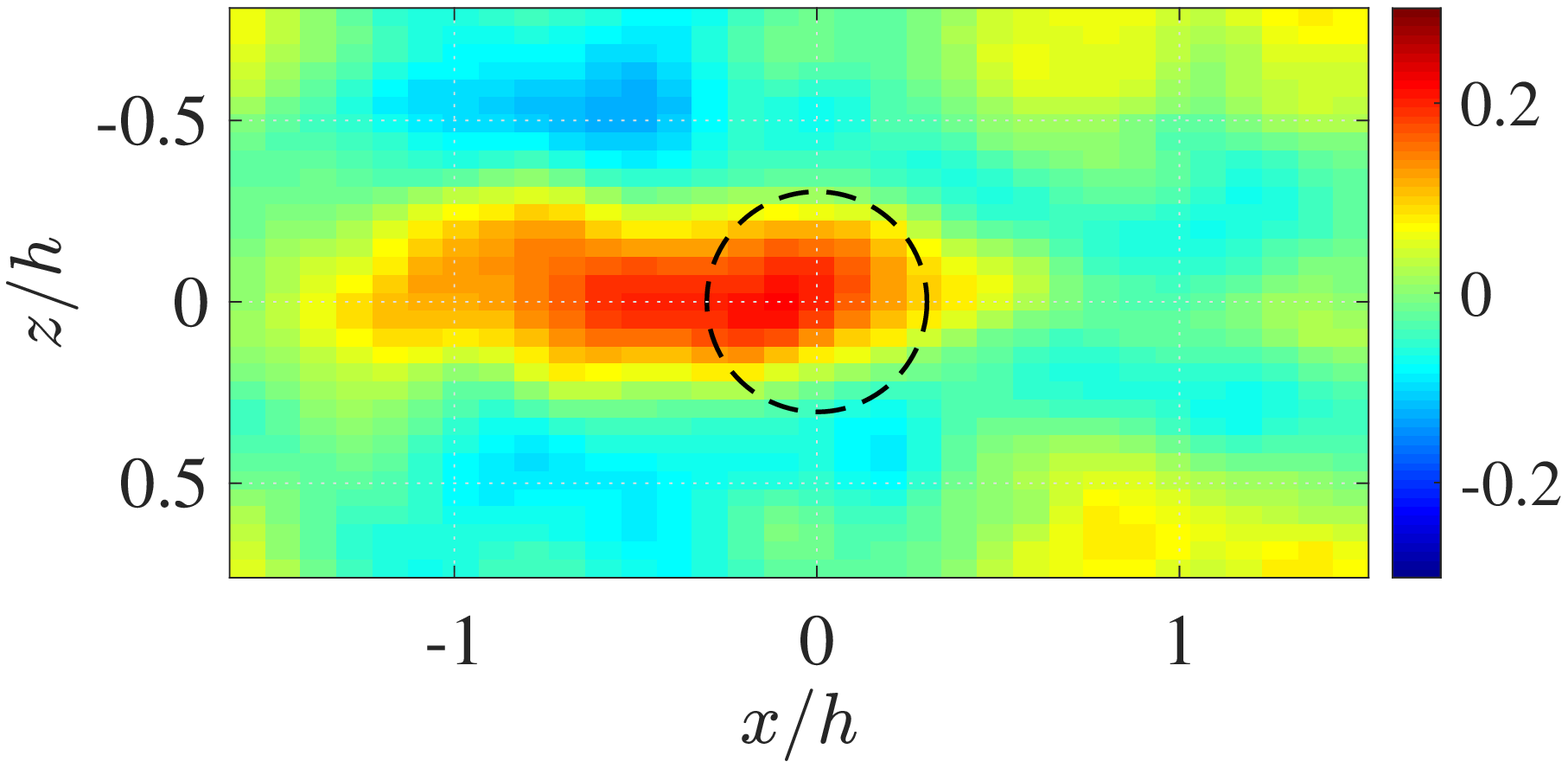}
\mylab{-5.3cm}{2.5cm}{$(b)$}
\hfill
\includegraphics[width=0.32\textwidth]{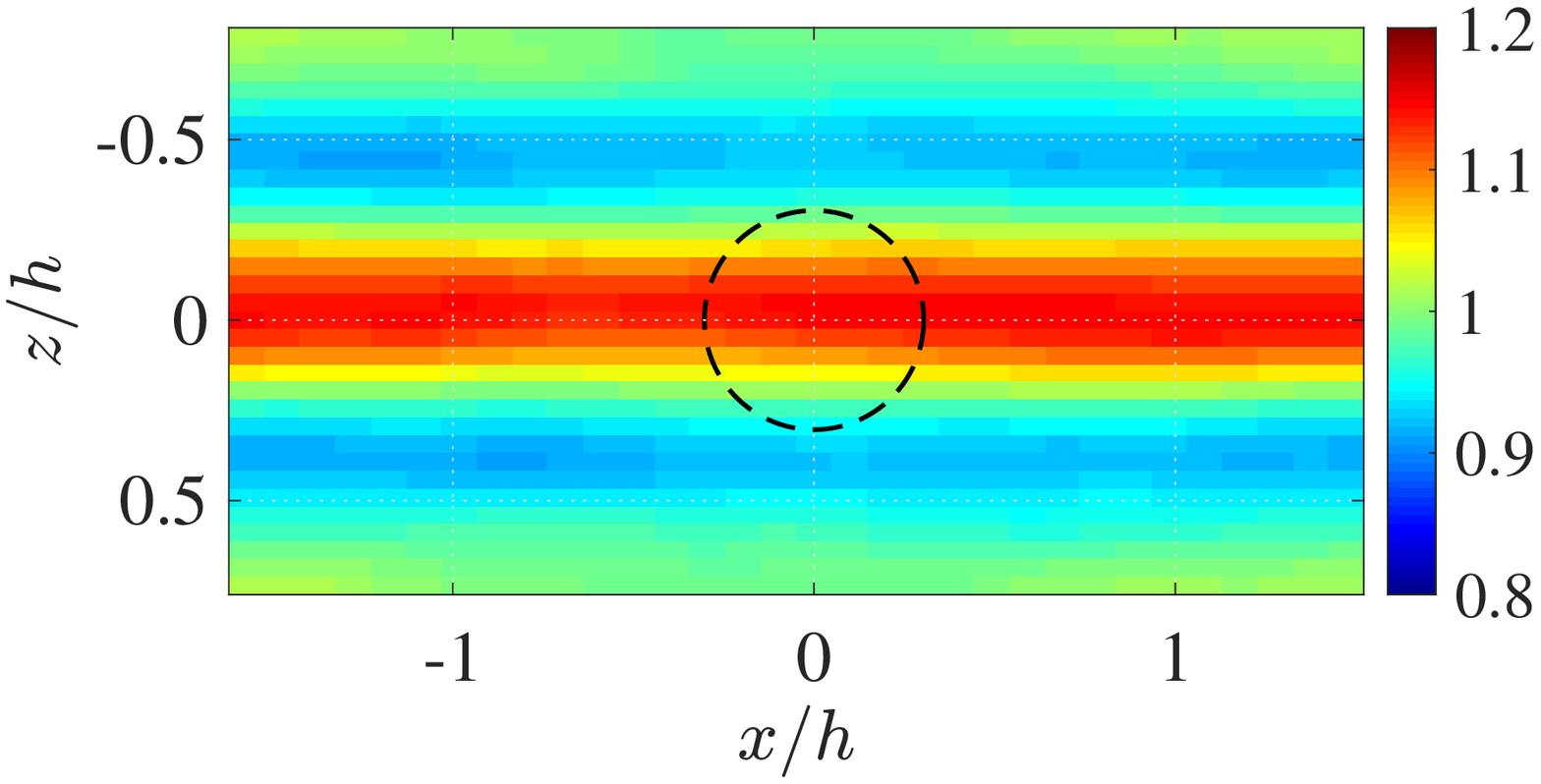}
\mylab{-5.3cm}{2.5cm}{$(c)$}
\vspace{1mm}

\includegraphics[width=0.32\textwidth]{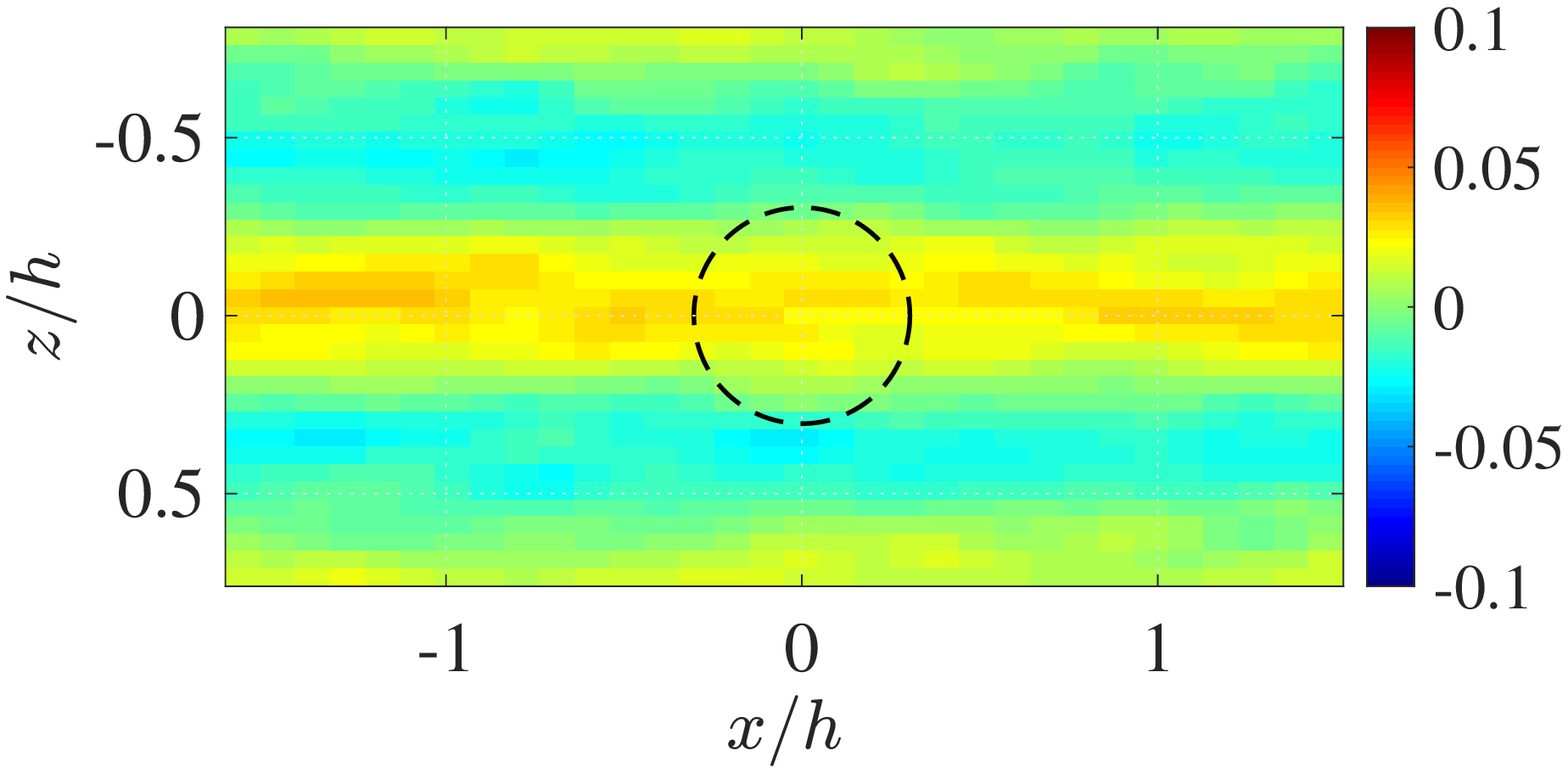}
\mylab{-5.3cm}{2.5cm}{$(d)$}
\hfill
\includegraphics[width=0.32\textwidth]{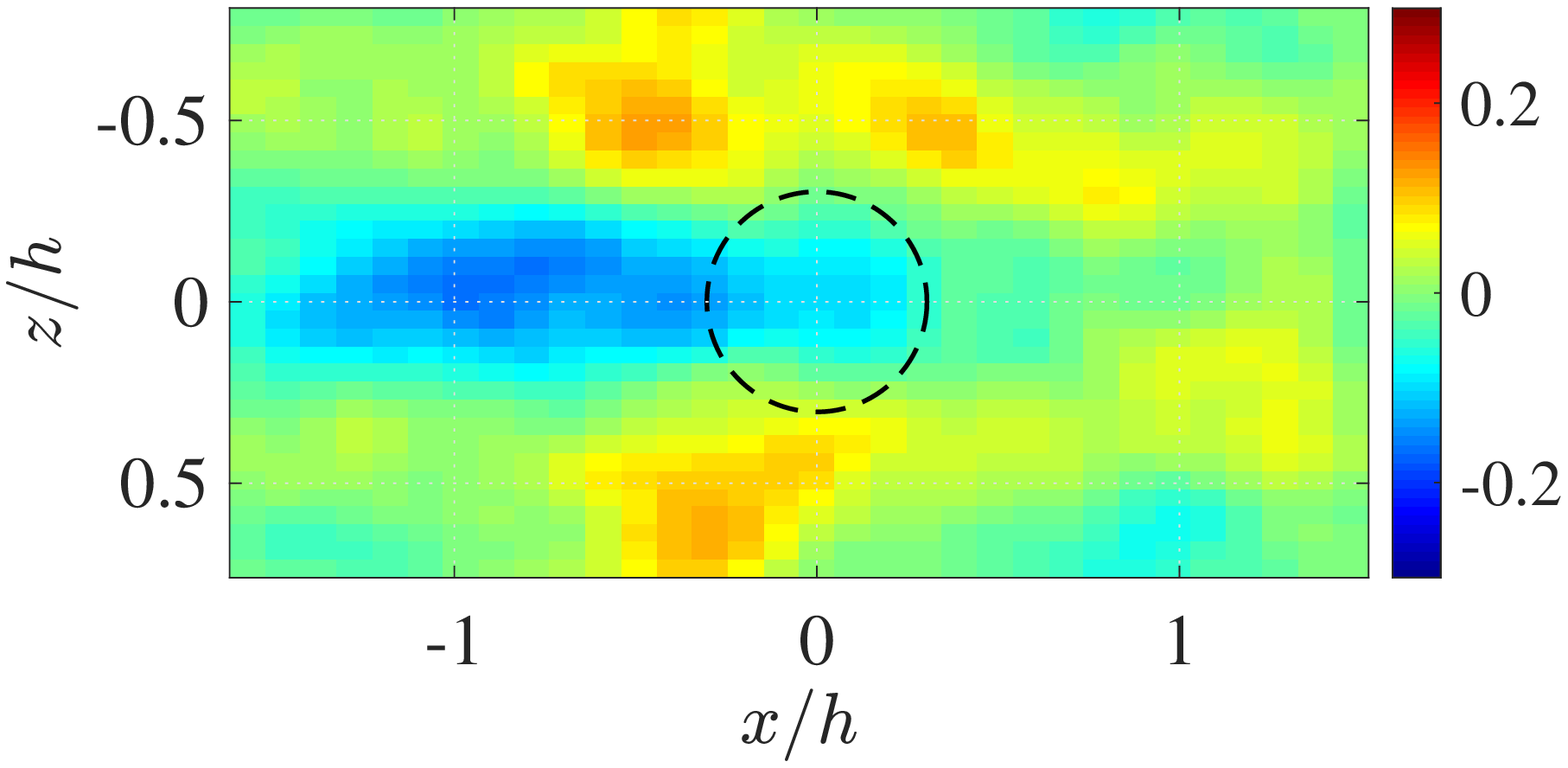}
\mylab{-5.3cm}{2.5cm}{$(e)$}
\hfill
\includegraphics[width=0.32\textwidth]{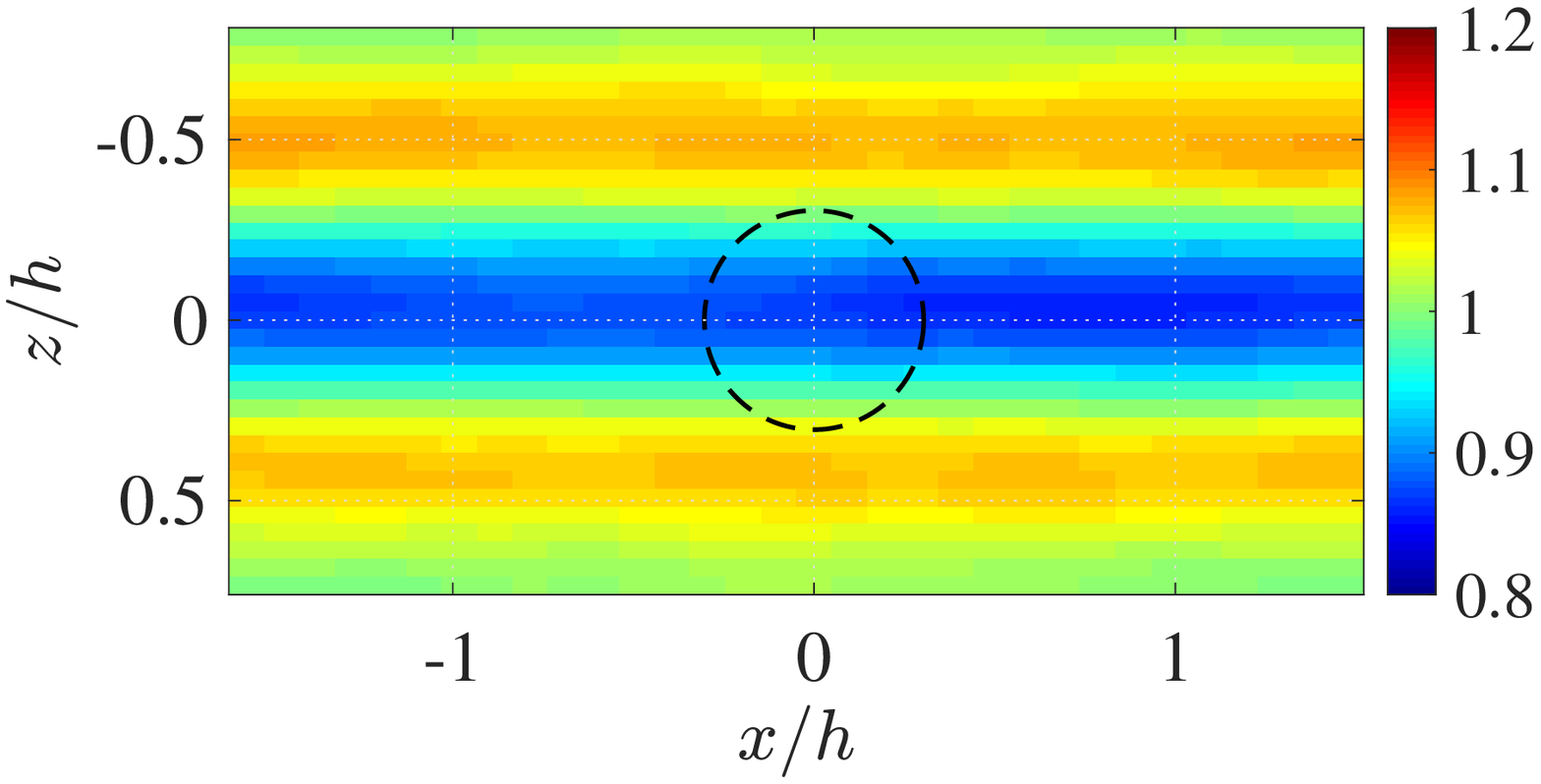}
\mylab{-5.3cm}{2.5cm}{$(f)$}
\caption{
\label{fig:state_c} 
Fluid state before actuation ($L_f^+ = 100$, $T_f^+ = 25$) conditioned to $\overline \Delta \tau (h/u_{\tau}) > 0.01 \tau_0$.     
Top row ($a,b,c$), up-ward forcing, $f_0>0$. 
Bottom row ($d,e,f$),  downward forcing, $f_0<0$. 
Left column ($a,d$), vertical velocity at $y^+=10$. 
Centre column ($b,e$), wall pressure. 
Right column ($c,f$), wall shear. 
Velocity, pressure and shear are normalised with $\tau_0$.  
The position and size of the actuator is indicated with the circle with dashed line.
}
\end{figure}

The states that correspond to the drag reduction shown in the CDFs are characterised by the conditionally averaged velocity, pressure and wall-stress fields shown in figure \ref{fig:state_c}. The figure shows the results conditioned to $\overline{\Delta \tau}(h/u_\tau) > 0.01 \tau_0$, for the positive and negative forcings with $L_f^+= 100$ and $T_f^+=25$. Pressure, velocity and wall-shear are normalised in wall units, using the friction velocity of the base case. 

The results suggests that, on average, the forcing operates as an opposition control when it produces drag reduction. The conditioned fields for $f_0>0$ show a Q2 event, characterised by a negative vertical velocity structure aligned with the position of the actuator, with an associated high velocity streak (as suggested by wall-shear stresses larger than $\tau_0$) and a high pressure region at the wall, slightly upstream of the actuator. On the other hand, drag reduction seems to be achieved by the negative forcing $f_0<0$ when acting on Q4 event: a positive vertical velocity structure within a low-velocity streak (i.e, $\mu\partial_y u|_{y=0} < \tau_0$), with an associated negative wall-pressure fluctuation slightly upstream of the actuator. It should be noted that, the intensity of the conditionally averaged $v$ and $p$ fields is relatively small, of the order of 1/3 of the standard deviation of $v$ and $p$ in the base flow. On the other hand, the intensity of the conditionally averaged wall-shear is comparable to the instantaneous fluctuations $\mu \partial_y u|_{y=0}$. 
This suggests that drag reduction is achieved by acting directly on strong wall-shear events, and not on the vertical motions that generate them.  

\subsection{{\em A posteriori} analysis: sensor definition and performance of the actuation triggered by the sensor.} 

Based on the results of the {\em a priori} analysis, a preliminary evaluation of a pair of sensor/actuator is presented next. The case considered for this analysis is also the forcing with $T_f^+=25$ and $L_f^+=100$, as in the previous section. Two different sensors are defined and analysed separately, using the same episodes used for the {\em a priori} analysis. One for the pressure at the wall, and one for the shear stresses at the wall. 
Each sensor measures the spatially averaged values of $p$ or $\mu\partial_y u$ in a circle of diameter $L_f$ ($p_s$ and $\tau_s$, respectively). The sensors are placed in the lower wall of the channel, 
at the same spanwise location as the actuator, at a streamwise coordinate $x_s$ measured with respect to the actuator. 

Based on the results shown on figure \ref{fig:state_c}, the actuation is triggered when the sensors exceed certain thresholds.  
For the pressure triggered cases, 
the upward forcing ($f_0>0$) is triggered by the pressure sensor when $p_s > 0.8 u_\tau^2$, and the downward forcing ($f_0<0$) is triggered when $p_s < -0.8 u_\tau^2$. 
For the wall-shear triggered cases, the upward and downward forcings are triggered by $\tau_s > 1.2\tau_0$ and $\tau_s < 0.8\tau_0$, respectively. For these thresholds, and using a sensor of size $L_f^+=100$, the actuator triggers roughly 15-20\% of the time, 
which is of the same order of magnitude as the probability of $\overline{\Delta \tau}(h/u_\tau) > 0.01\tau_0$ for the same actuation. 

\begin{figure}
\includegraphics[width=0.47\textwidth]{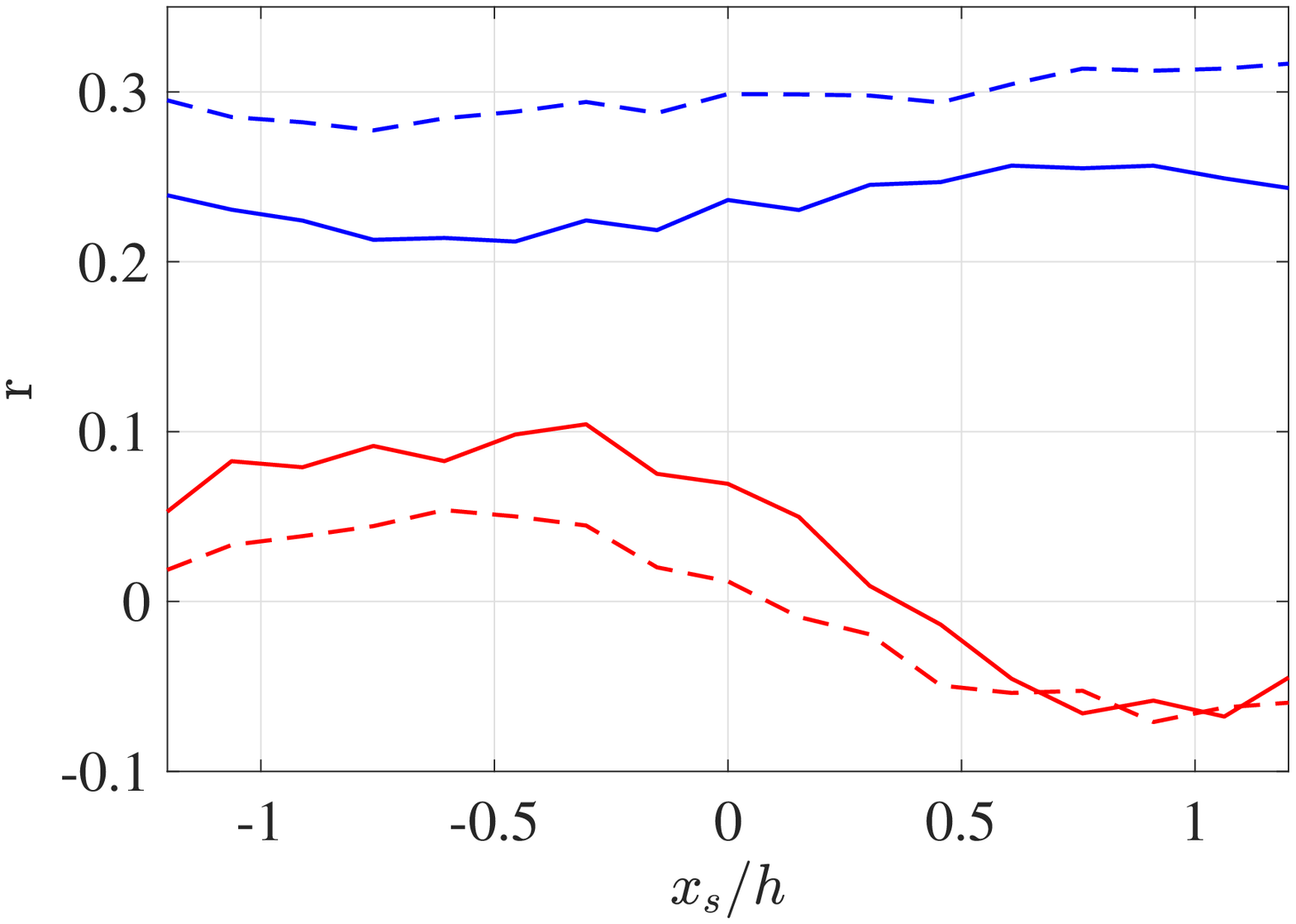}
\mylab{-6.4cm}{4.9cm}{$(a)$}
\hfill
\includegraphics[width=0.49\textwidth]{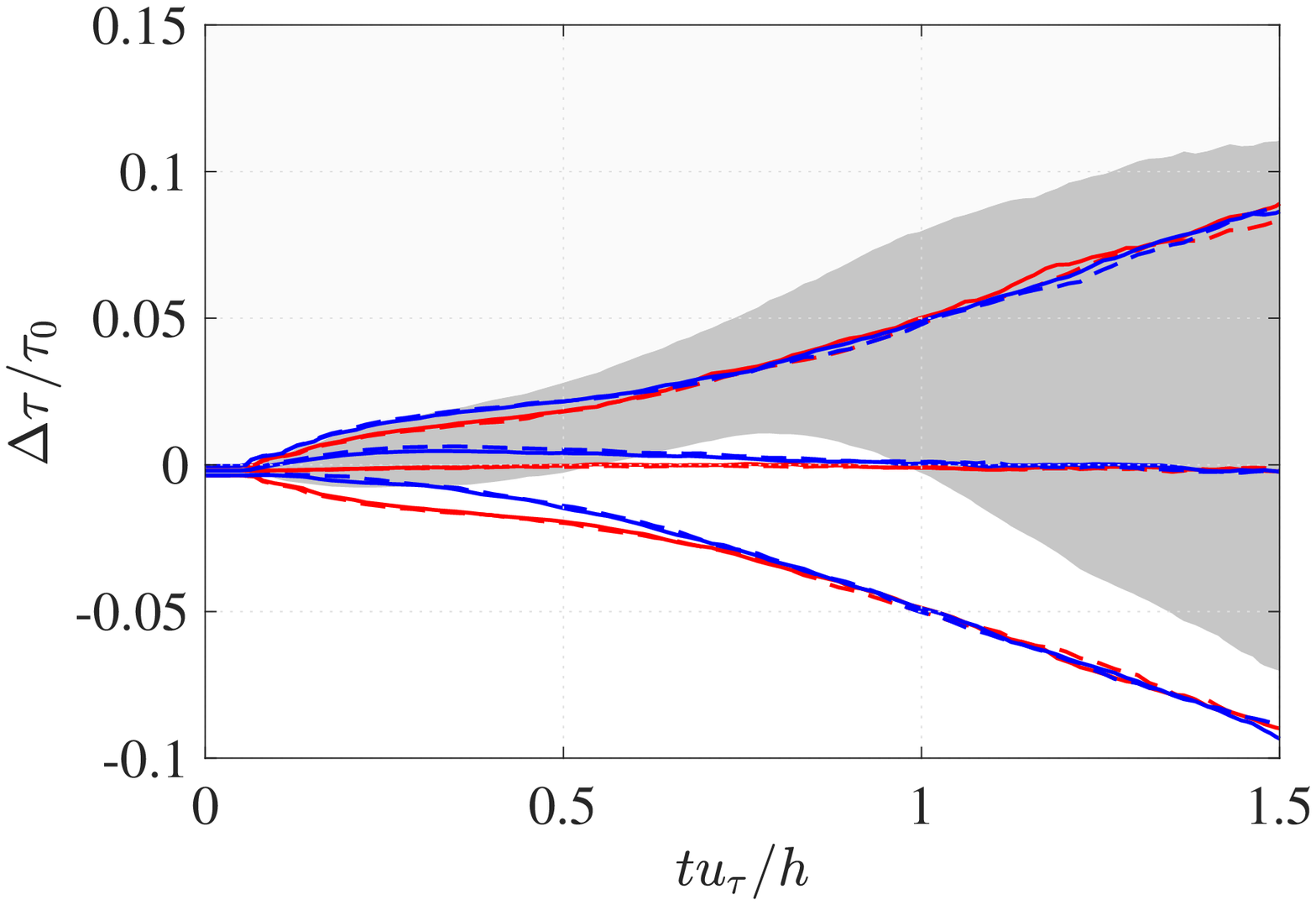}
\mylab{-6.4cm}{4.9cm}{$(b)$}
\caption{
\label{fig:aposteriori} 
($a$) Ratio between the mean value of $\Delta \tau$ conditioned to sensors and conditioned to $\overline{\Delta \tau}$, as defined in equation (\ref{eq:merit}). 
($b$) Cumulative Distribution Function (CDF) of $\Delta \tau$ conditioned to sensors located at $x_s = -0.5h$,  pressure sensor  in red and wall-shear sensor in blue. 
The shaded contour correspond to the CDF conditioned to $\overline{\Delta \tau}$, as show in figure \ref{fig:cdf_c}$b$. 
Solid lines for upward forcing ($f_0>0$), and dashed lines for downward forcing ($f_0<0$). 
Both panes correspond to case $T_f^+=25$ and $L_f^+ = 100$. 
}
\end{figure}

Figure \ref{fig:aposteriori}$a$ shows the variation of the performance of each pair of actuator/sensor with the location of the sensor, $x_s$. The selected figure of merit is 
\begin{equation} 
\label{eq:merit}
r = \frac{\langle \overline{\Delta \tau}(h/u_\tau) | \mbox{sensor} \rangle} { \langle \overline{\Delta \tau}(h/u_\tau) |  \overline{\Delta \tau}(h/u_\tau) > 0.01\tau_0  \rangle},  
\end{equation} 
which corresponds to the ratio between the mean value of $\Delta \tau$ for $t<h/u_\tau$ when the sensor triggers the actuator, and the mean value of the drag reduction obtained in the {\em a priori} analysis. In other words, $r$ measures to what extent the sensor-based-actuation is able to replicate the results of the {\em a priori} analysis. 

Figure \ref{fig:aposteriori}$a$ shows that the sensor based on $\tau$ has a better performance than the sensor based on $p_s$, at least in terms of the figure of merit defined in equation (\ref{eq:merit}).
The performance of the cases triggered by the wall-shear varies little with the position of the sensor, while the performance of the cases triggered by the pressure sensor shows a stronger dependency on  $x_s$. 
This result is consistent with the conditionally averaged pressure and wall-shear fields obtained in the {\em a priori} analysis (figure \ref{fig:state_c}). It is also interesting to note that, when the trigger is pressure based, the upward forcing seems to perform slightly better than the downward forcing, while the oposite is true when the actuation is triggered by the wall-shear. 

In any case, the drag reduction of the pairs sensor/actuator is significantly less than that observed in the {\em a priori} analysis. Part of the problem seems to be that the sensor/actuator pairs fail to achieve a long lasting effect on the flow observed in the {\em a prioir} analysis
($r\lesssim0.3$ for the wall-shear triggered actuation, and $r\lesssim0.1$ for the pressure triggered actuation). This is shown in figure \ref{fig:aposteriori}$b$, where the CDF of $\Delta \tau$ conditioned to trigger by sensor (lines) and the CDF of $\Delta \tau$ conditioned to $\overline{\Delta \tau}(h/u_\tau)>0.01\tau_0$ (shaded contour) are compared. 
Only the wall-shear triggered cases show a shift of the CDF to positive values for short times, while the CDF for the pressure based sensor is roughly symmetrical. Overall, the comparison of the CDF of the {\em a priori} and {\em a posteriori} analysis suggest that more work is needed to design efficient forcing strategies (maybe, combining pressure and wall-shear sensors).

\section{Conclusions} 

A Monte Carlo experiment has been presented to evaluate how localised actuation and sensing limits our ability to control skin friction in turbulent flow. The study is mostly a proof of concept, and as such it uses a toy problem for the turbulent flow: a channel flow at a friction Reynolds number $Re_\tau = 165$ in a small computational domain.  The simulations are run with a GPU version of the channel code, resulting in a considerable speed-up of the simulations. Twelve different actuations are considered, with positive (upward) and negative (downward) forcings, two sizes ($L_f^+ = 50$ and $100$) and three different durations ($T_f^+ = 25, 50$ and $100$).

The evaluation of the statistics of the differences in the skin friction of the base case and the forced cases shows that the probability of a given forcing to increase or decrease the skin friction is roughly the same. 
As expected, increasing the size or the duration of the forcing increases its effect on the flow.
Forcings with durations longer than a flow-through time (i.e., the time that a fluid particle takes to cross the streamwise length of the computational domain) show a clear tendency to increase skin friction. The effects of all forcings are forgotten quickly in the CDF of $\Delta \tau$, and after $2 - 3 h/u_\tau$ the flow recovers a statistical state comparable to the unforced channel. 

Two different analysis are performed with the database generated in the Monte Carlo experiment. In the {\em a priori} analysis, the initial conditions that result in drag reduction after the actuation are investigated. It is observed that drag reduction is obtained when the actuation opposes Q2 or Q4 events, identified here by the vertical velocity at $y^+=10$ and the shear-stresses at the wall. These events also leave a footprint in the pressure at the wall, with a positive (negative) pressure fluctuation associated to a Q4 (Q2) event. 
The drag reduction obtained in the {\em a priori} analysis is small. For instance, for the case with $L_f^+ = 100$ and $T_f^+=25$, drag reducing conditions are observed around 25\% of the time, and the averaged drag reduction after the actuation is about 3-4\%. 
This limited drag reduction is not surprising, since the actuation only takes place in a small percentage of the area of the bottom wall of the channel (i.e., 7\% for the cases with $L_f^+ = 100$, less than 2\% for the cases with $L_f^+=50$). 

After the {\em a prioir} analysis, a preliminary design of a pair of sensor/actuator is attempted. using pressure and wall-shear sensors to trigger the actuation.  The analysis shows that the performance of the sensor/actuator is worse than the performance obtained in the {\em a priori} analysis, apparently because the sensors defined here are not able to identify those episodes where the effect of the actuator on the skin friction is long lasting. The performance of the sensor/actuator pair is slightly better for wall-shear sensors than for pressure sensors,  in agreement with the pressure and wall-shear fields conditioned to drag reduction computed in the {\em a priori} analysis. 
This suggests that, at least with the present set up, acting directly on relatively strong events that produce large fluctuations of $\tau$ seems more efficient than trying to prevent the formation of these strong events. 

Overall, the Monte Carlo experiment described here has proven to be an interesting tool to interrogate the flow. Its value for develop control strategies for wall-bounded turbulent flows should be proved in subsequent studies.  \\[1ex]

%

\noindent{\bf Acknowledgments: } This work has been supported by the European Research Council, under the COTURB grant ERC-2014.AdG-669505.
The authors are grateful to M. Encinar and  A. Guseva for fruitful discussions during the Madrid Summer Programme, 2019.  

\bibliographystyle{plainnat}
\bibliography{bibs}

\end{document}